\documentclass[12pt]{article}
\usepackage{makeidx}
\usepackage{amsfonts}
\usepackage{amssymb}
\usepackage{amsmath}

\newtheorem{definition}{Definition}

\newtheorem{lemma}{Lemma}
\newtheorem{proposition}{Proposition}

\begin{document}

\title{On the Choi-Jamiolkowski Correspondence \\
in Infinite Dimensions\footnote{Support by the Institute Mittag-Leffler (Djursholm, Sweden) is gratefully acknowledged.}}
\author{A. S. Holevo \\
Steklov Mathematical Institute}
\date{}
\maketitle

\begin{abstract}
We give a mathematical formulation for the Choi-Jamiolkowski (CJ)
correspondence in the infinite-dimensional case close to one used in quantum
information theory. We show that ``unnormalized maximally entangled state''
and the corresponding analog of the Choi matrix can be defined rigorously as
positive semidefinite forms on an appropriate dense subspace. The properties
of these forms are discussed in Sec. 2. In Sec. 3 we prove a version of a
result from \cite{oppsw} characterizing the form corresponding to
entanglement-breaking channel by giving precise definitions of the separable
CJ form and the relevant integral. In Sec. 4,5 we obtain explicit
expressions for CJ forms and operators defining Bosonic Gaussian channel. In
particular, a condition for existence of the bounded CJ operator is given.
\end{abstract}

\section{Introduction}

In this paper we give a mathematical formulation for the
Choi-Jamiolkowski (CJ) correspondence (\cite{jam}, \cite{choi}) in
the infinite-dimensional case in the form close to one used in
quantum information theory (see e.g. \cite{nc}). We show that there
is no need to use a limiting procedure (cf. \cite{gc}) to define
``unnormalized maximally entangled state'' and the corresponding
analog of the Choi matrix \cite{choi} since they can be defined
rigorously as, in general, nonclosable forms on an appropriate dense
subspace. The properties of these forms are discussed in Sec. 2. An
important question is: when the CJ form is given by a bounded
operator. This is the case for entanglement-breaking channels: we
prove this in Sec. 3 along with a version of a result from
\cite{oppsw} characterizing CJ operators which correspond to such
channels by giving precise definitions of a separable operator and a
relevant integral. In Sec. 4 we obtain explicit expressions for CJ
forms and operators defining a general Bosonic Gaussian channel. In
Sec. 5 we give a decomposition of CJ form into product of the four
principal types and a necessary and sufficient condition for
existence of the bounded CJ operator.

\section{Positive semidefinite forms}

In what follows $\mathcal{H},\mathcal{K},\dots $ denote separable Hilbert
spaces; $\mathfrak{T}(\mathcal{H})$ denotes the Banach space of trace-class
operators in $\mathcal{H}$, and $\mathfrak{S}(\mathcal{H})$ -- the convex
subset of all density operators. We shall also call them \textit{states }for
brevity, having in mind that a density operator $\rho $ uniquely determines
a normal state on the algebra $\mathfrak{B}(\mathcal{H})$ of all bounded
operators in $\mathcal{H}$. Equipped with the trace-norm distance, $
\mathfrak{S}(\mathcal{H})$ is a complete separable metric space. It is known
\cite{D-A}, \cite{d2} that a sequence of quantum states $\left\{ \rho
_{n}\right\} $ converging to a state $\rho $ in the weak operator topology
converges to it in the trace norm. Moreover, it suffices that $
\lim_{n}\langle \psi |\rho _{n}|\psi \rangle =\langle \psi |\rho |\psi
\rangle $ for $\psi $ in a dense linear subspace of $\mathcal{H}$ .

\begin{definition}
A \textit{channel} is a linear map $\Phi $:$\mathfrak{T}( \mathcal{H}
_{A})\mapsto \mathfrak{T}(\mathcal{H}_{B})$ with the properties:

1) $\Phi (\mathfrak{S}(\mathcal{H}_{A}))\subseteq \mathfrak{S}(\mathcal{H}
_{B});$ this implies that $\Phi $ is bounded map \cite{d2} and hence it is
uniquely determined by the infinite matrix $\left[ \Phi \left( |i\rangle
\langle j|\right) \right] ,$ where $\left\{ |i\rangle \right\} $ is a fixed
orthonormal basis in $\mathcal{H}_{A}.$

2) The block matrix $\left[ \Phi \left( |i\rangle \langle j|\right) \right] $
is positive semidefinite in the sense that for any finite collection of
vectors $\left\{ |\psi _{i}\rangle \right\} \subseteq \mathcal{H}_{B}$
\begin{equation}
\sum_{ij}\langle \psi _{i}|\Phi \left( |i\rangle \langle j|\right) |\psi
_{j}\rangle \geq 0.  \label{cp}
\end{equation}

\bigskip The \textit{Choi-Jamiolkowski (CJ) form} of the channel, associated
with the basis $\left\{ |i\rangle \right\} ,$ is defined by the relation (\ref{relation}) below.
\end{definition}

In $\mathcal{H}_{B}\otimes \mathcal{H}_{A}$ we consider the dense domain
which is invariant under \textquotedblleft local\textquotedblright\ bounded
operators $X_{B}\otimes X_{A}:$
\begin{equation*}
\mathcal{H}_{B}\times \mathcal{H}_{A}=\mathrm{lin}\left\{ \psi _{B}\otimes
\psi _{A}:\psi _{B}\in \mathcal{H}_{B},\psi _{A}\in \mathcal{H}_{A}\right\} .
\end{equation*}

\begin{lemma}
There is a unique sesquilinear positive semidefinite form $\Omega _{\Phi }$
on $\mathcal{H }_{B}\times \mathcal{H}_{A}$ satisfying the relation
\begin{equation}
\Omega _{\Phi }\left( \psi _{B}\otimes \psi _{A};\psi _{B}^{\prime }\otimes
\psi _{A}^{\prime }\right) =\langle \psi _{B}|\Phi \left( |\bar{\psi}
_{A}\rangle \langle \bar{\psi}_{A}^{\prime }|\right) |\psi _{B}^{\prime
}\rangle =\langle \bar{\psi}_{A}^{\prime }|\Phi ^{\ast }\left( |\psi
_{B}^{\prime }\rangle \langle \psi _{B}|\right) |\bar{\psi}_{A}\rangle ,
\label{relation}
\end{equation}
where $|\bar{\psi}\rangle =\sum_{i=1}^{+\infty }|i\rangle \overline{\langle
i|\psi \rangle }.$
\end{lemma}

\textit{Proof.} It is sufficient to show that for any $|\psi _{BA}\rangle
=\sum\limits_{j}|\psi _{B}^{j}\rangle \otimes |\psi _{A}^{j}\rangle \in
\mathcal{H}_{B}\times \mathcal{H}_{A}$
\begin{equation*}
\sum\limits_{jk}\langle \psi _{B}^{j}|\Phi \left( |\bar{\psi}
_{A}^{j}\rangle \langle \bar{\psi}_{A}^{k}|\right) |\psi _{B}^{k}\rangle
\geq 0,
\end{equation*}
and $|\psi _{BA}\rangle =0$ implies that the above sum is equal to zero.
Decomposing $|\psi _{A}^{j}\rangle =\sum_{i=1}^{+\infty }c_{ij}|i\rangle ,$
we have $|\psi _{BA}\rangle =\sum\limits_{i}|\tilde{\psi}_{B}^{i}\rangle
\otimes |i\rangle $ where $|\tilde{\psi}_{B}^{i}\rangle
=\sum\limits_{j}c_{ij}|\psi _{B}^{j}\rangle $ and the above sum is equal to $
\sum\limits_{ij}\langle \tilde{\psi}_{B}^{i}|\Phi \left( |i\rangle \langle
j|\right) |\tilde{\psi}_{B}^{j}\rangle $ whence the result follows. $\square$

Note that for any orthonormal basis $\left\{ |e_{k}\rangle \right\} $ in $
\mathcal{H}_{B}$
\begin{equation}
\sum_{k}\Omega _{\Phi }\left( e_{k}\otimes \psi _{A};e_{k}\otimes \psi
_{A}^{\prime }\right) =\langle \psi _{A}|\psi _{A}^{\prime }\rangle .
\label{partial}
\end{equation}
The expression on the left is a natural form generalization of partial trace
with respect to $\mathcal{H}_{B}$. The relation (\ref{partial}) shows that
for $\Omega _{\Phi }$ it is always given by the (form corresponding to) the
identity operator $I_{A}$. Similarly defined partial trace with respect to $
\mathcal{H}_{A}$ not always exists; however this is the case when the
expression $\Phi \lbrack I_{A}]$ is well-defined in the sense of \cite{h1}
and then it is given by that operator. Positivity and the property (\ref
{partial}) imply
\begin{equation}
\left\vert \Omega _{\Phi }\left( \psi _{B}\otimes \psi _{A};\psi
_{B}^{\prime }\otimes \psi _{A}^{\prime }\right) \right\vert \leq \left[
\Omega _{\Phi }\left( \psi _{B}\otimes \psi _{A};\psi _{B}\otimes \psi
_{A}\right) \Omega _{\Phi }\left( \psi _{B}^{\prime }\otimes \psi
_{A}^{\prime };\psi _{B}^{\prime }\otimes \psi _{A}^{\prime }\right) \right]
^{1/2}  \notag
\end{equation}
\begin{equation}
\leq \left\Vert \psi _{B}\right\Vert \left\Vert \psi _{B}^{\prime
}\right\Vert \left\Vert \psi _{A}\right\Vert \left\Vert \psi _{A}^{\prime
}\right\Vert .  \label{bound}
\end{equation}

Conversely, any sesquilinear positive semidefinite form $\Omega $ on $
\mathcal{H}_{B}\times \mathcal{H}_{A}$ satisfying the condition (\ref
{partial}) uniquely defines a channel $\Phi $ such that $\Omega _{\Phi
}=\Omega $ (via the reversed relation (\ref{relation})). The positivity of
the form $\Omega _{\Phi }$ can be used to prove the Kraus decomposition for $
\Phi $ similarly to the finite-dimensional case \cite{nc}. Indeed, let $
\mathfrak{H}$ be the Hilbert space obtained by completion of $\mathcal{H}
_{B}\times \mathcal{H}_{A}$ with respect to the inner product defined by the
(factorized) form $\Omega _{\Phi }.$ Then any orthonormal basis in $
\mathfrak{H}$ defines a countable collection of linear functionals $
\{|f_{l}\rangle \}_{l=1}^{+\infty }$ on $\mathcal{H}_{B}\times \mathcal{H}
_{A}$ such that
\begin{equation*}
\Omega _{\Phi }\left( \psi _{B}\otimes \psi _{A};\psi _{B}^{\prime }\otimes
\psi _{A}^{\prime }\right) =\sum_{l=1}^{+\infty }f_{l}\left( \psi
_{B}^{\prime }\otimes \psi _{A}^{\prime }\right) \overline{f_{l}\left( \psi
_{B}\otimes \psi _{A}\right) }.
\end{equation*}
Define the linear operators $V_{l}:\mathcal{H}_{A}\rightarrow \mathcal{H}
_{B} $ by the relation $\langle \psi _{B}|V_{l}|\psi _{A}\rangle =\overline{
f_{l}\left( \psi _{B}\otimes \bar{\psi}_{A}\right) },$ then (\ref{partial})
implies that $V_{l}$ are bounded operators satisfying \newline
$\sum_{l=1}^{+\infty }V_{l}^{\ast }V_{l}=I_{A}$ and (\ref{relation}) implies
the Kraus decomposition $\Phi \left( \rho \right) =\sum_{l=1}^{+\infty
}V_{l}\rho V_{l}^{\ast }.$

If the form $\Omega _{\Phi }$ is closable \cite{reed}, then it is defined by
the unique densely defined selfadjoint positive operator, which we also
denote $\Omega _{\Phi }.$ If the domain of the form $\Omega _{\Phi }$ is the
whole $\mathcal{H}_{B}\otimes \mathcal{H}_{A}$ then the operator $\Omega
_{\Phi }$ is bounded. The property (\ref{partial}) then reads
\begin{equation}
\mathrm{Tr}_{B}\Omega _{\Phi }=I_{A}.  \label{partialop}
\end{equation}
In this case the defining relation (\ref{relation}) can be given a more familiar
form
\begin{equation*}
\mathrm{Tr}\Omega _{\Phi }\left( \rho \otimes X\right) =\mathrm{Tr}\Phi
\left( \rho ^{\top }\right) X=\mathrm{Tr}\rho ^{\top }\Phi ^{\ast }\left(
X\right) ,\quad \rho \in \mathfrak{T}(\mathcal{H}),X\in \mathfrak{B}(
\mathcal{H}),
\end{equation*}
where $^{\top}$ denotes transposition in the basis $\left\{
|i\rangle \right\}$, so that $|\bar{\psi} _{A}\rangle \langle
\bar{\psi}_{A}^{\prime }|=\left( |\psi _{A}^{\prime }\rangle \langle
\psi_{A}|\right)^{\top}$.

An example of nonclosable sesquilinear form is provided by the identity
channel $\Phi =\mathrm{Id}$, for which
\begin{equation*}
\Omega _{\mathrm{Id}}\left( \psi _{B}\otimes \psi _{A};\psi _{B}^{\prime
}\otimes \psi _{A}^{\prime }\right) =\langle \psi _{B}\otimes \psi
_{A}|\Omega \rangle \langle \Omega |\psi _{B}^{\prime }\otimes \psi
_{A}^{\prime }\rangle ,
\end{equation*}
where $\langle \Omega |$ is the unbounded linear form on $\mathcal{H}
_{A}\times \mathcal{H}_{A}$ defined as
\begin{equation*}
\langle \Omega |\psi _{1}\otimes \psi _{2}\rangle =\sum_{i=1}^{+\infty
}\langle i|\psi _{1}\rangle \langle i|\psi _{2}\rangle ;\quad \psi _{1},\psi
_{2}\in \mathcal{H}_{A},
\end{equation*}
and $|\Omega \rangle $ is the dual antilinear form which represents
\textquotedblleft unnormalized maximally entangled state\textquotedblright ,
$|\Omega \rangle =\sum_{i=1}^{+\infty }|i\rangle \otimes |i\rangle .$ The
relation
\begin{equation*}
\Omega _{\Phi }=\left( \Phi \otimes \mathrm{Id}_{A}\right) \left( \Omega _{
\mathrm{Id}}\right)
\end{equation*}
holds in the weak sense i.e. as equality for the forms defined on $\mathcal{
\ \ H}_{B}\times \mathcal{H}_{A}.$

Notice that $\left( X_{A}\otimes X_{B}\right) |\Omega \rangle
=\left( I\otimes X_{B}X_{A}^{\top }\right) |\Omega \rangle $.  If $
\left\{ |i\rangle ^{\prime }\right\} $ is another basis such that
$|i\rangle ^{\prime }=U|i\rangle $ for a unitary $U,$ then $|\Omega
\rangle ^{\prime }=\left( I\otimes UU^{\top }\right) |\Omega \rangle
$ and
\begin{equation*}
\Omega _{\Phi }^{\prime }=\left( I\otimes UU^{\top }\right) \Omega _{\Phi
}\left( I\otimes UU^{\top }\right) ^{\ast }.
\end{equation*}
In the case of bounded $\Omega _{\Phi }$ this implies that $\left\Vert
\Omega _{\Phi }\right\Vert $ is the same for all choices of the basis $
\left\{ |i\rangle \right\} .$ Notice also that transposition in the basis $
\left\{ |i\rangle ^{\prime }\right\} $ is given by $X^{\top ^{\prime
}}=\left( UU^{\top }\right) X^{\top }\left( UU^{\top }\right) ^{\ast }.$

Fix a state $\sigma $ in $\mathfrak{S}(\mathcal{H}_{A})$ of full rank, and
let $\{|i\rangle \}_{i=1}^{+\infty }$ be the basis of eigenvectors of $
\sigma $ with the corresponding (positive) eigenvalues $\{\lambda
_{i}\}_{i=1}^{+\infty }$. Consider the\ purifying vector
\begin{equation*}
|\psi _{\sigma }\rangle =\sum_{i=1}^{+\infty }\lambda _{i}^{1/2}|i\rangle
\otimes |i\rangle
\end{equation*}
in the space $\mathcal{H}_{A}\otimes \mathcal{H}_{A}$. Then the state
\begin{equation}
\rho _{\Phi }(\sigma )=(\Phi \otimes \mathrm{Id}_{A})(|\psi _{\sigma
}\rangle \langle \psi _{\sigma }|)\in \mathfrak{S}(\mathcal{H}_{B}\otimes
\mathcal{H}_{A})
\end{equation}
satisfying $\mathrm{Tr}_{B}\rho _{\Phi }(\sigma )=\sigma $ uniquely
determines the channel $\Phi $ via the relation

\begin{equation*}
\Phi (|i\rangle \langle j|)=\lambda _{i}^{-1/2}\lambda _{j}^{-1/2}\mathrm{Tr}
_{A}(I_{B}\otimes |j\rangle \langle i|)\rho _{\Phi }(\sigma ),
\end{equation*}
see \cite{hsw}. The connection between $\rho _{\Phi }(\sigma )$ and $\Omega
_{\Phi }$ is
\begin{equation*}
\langle \psi _{B}\otimes \psi _{A}|\rho _{\Phi }(\sigma )|\psi _{B}^{\prime
}\otimes \psi _{A}^{\prime }\rangle =\Omega _{\Phi }\left( \psi _{B}\otimes
\sigma ^{1/2}\psi _{A};\psi _{B}^{\prime }\otimes \sigma ^{1/2}\psi
_{A}^{\prime }\right) .
\end{equation*}
Note that $\Omega _{\Phi }$ is uniquely defined by its values on the dense
domain $\mathcal{D}=\mathcal{H}_{B}\times \sigma ^{1/2}(\mathcal{H}_{A})$
due to the property (\ref{bound})$.$

\section{Separable operators and entanglement-breaking channels}

Let $\sigma $ be a state in $\mathfrak{S}(\mathcal{H}_{A})$ of full rank and
$\Omega $ is a bounded positive operator satisfying (\ref{partialop}), then $
\sigma _{BA}=\left( I_{B}\otimes \sigma ^{1/2}\right) \Omega \left(
I_{B}\otimes \sigma ^{1/2}\right) $ is a density operator in $\mathcal{H}
_{B}\otimes \mathcal{H}_{A}$ such that $\mathrm{Tr}_{B}\sigma _{BA}=\sigma .$
Let us remind that a state $\rho \in \mathfrak{S}(\mathcal{H}_{B}\otimes
\mathcal{H}_{A})$\ is called\textit{\ separable }if it is in the convex
closure (in the weak operator topology and hence in the trace norm) of the
set of all product states. Separable states are precisely those which admit
the representation
\begin{equation}
\rho =\int_{\mathcal{X}}\left( \rho _{B}(x)\otimes \rho _{A}(x)\right) \mu
(dx),  \label{sepa}
\end{equation}
where $\mu $ is a Borel probability measure on a complete separable metric
space $\mathcal{X}$, see \cite{hsw}.

A channel $\Phi $ is called \textit{entanglement-breaking } if for arbitrary
state $\omega \in \mathfrak{S}(\mathcal{H}_{A}\otimes \mathcal{H}_{A})$ the
state $(\Phi \otimes \mathrm{Id}_{A})(\omega )$ is separable. Channel $\Phi $
is entanglement-breaking if and only if there is a complete separable metric
space $\mathcal{X}$, a Borel $\mathfrak{S}(\mathcal{H}_{B})$ -valued
function $x\mapsto \rho _{B}(x)$ and a probability operator-valued Borel
measure (POVM) $M_{A}(dx)$ on $\mathcal{X}$ in $\mathcal{H}_{A}$ such that
\begin{equation}
\Phi (\rho )=\int\limits_{\mathcal{X}}\rho _{B}(x)\mu _{\rho }(dx),
\label{ebr}
\end{equation}
where $\mu _{\rho }(E)=\mathrm{Tr}\rho M_{A}(E)$ for all Borel $E\subseteq
\mathcal{X}$. If $\rho =\rho _{\Phi }(\sigma ),$ then $\rho _{B}(x)$ in (\ref
{ebr}) is the same as in (\ref{sepa}) while $M_{A}(dx)$ is defined by the
relation
\begin{equation*}
\langle \psi _{A}^{\prime }|M_{A}(E)|\psi _{A}\rangle =\int_{E}\langle
\sigma ^{-1/2}\psi _{A}^{\prime }|\overline{\rho _{A}(x)}|\sigma ^{-1/2}\psi
_{A}\rangle \mu (dx);\quad \psi _{A},\psi _{A}^{\prime }\in \sigma
^{1/2}\left( \mathcal{H}_{A}\right) ,
\end{equation*}
where the complex conjugate is in the basis $\{|i\rangle \}.$

\begin{definition}
\label{sep} A bounded positive operator $\Omega $ in $\mathcal{H}_{B}\otimes
\mathcal{H}_{A}$ satisfying (\ref{partialop}) is called \textit{separable}
if it belongs to the closure, in the weak operator topology, of the convex
set of operators of the form $\sum\limits_{\alpha }\rho _{\alpha }\otimes
M_{\alpha },$ where $\left\{ M_{\alpha }\right\} $ is a finite resolution of
the identity in $\mathcal{H}_{A}$ and $\rho _{\alpha }\in \mathfrak{S} (
\mathcal{H}_{B}).$ The weakly closed convex set of separable operators will
be denoted $\mathfrak{C}_{BA}.$
\end{definition}

\begin{lemma}
For $\Omega \in \mathfrak{C}_{BA}$ the operator norm $\left\Vert \Omega
\right\Vert \leq 1.$
\end{lemma}

\textit{Proof.} It is sufficient to prove it for $\Omega
=\sum\limits_{\alpha }\rho _{\alpha }\otimes M_{\alpha },$ since the weak
operator limit does not increase the norm. Then
\begin{equation*}
\left\Vert \Omega \right\Vert =\sup_{\left\Vert \psi \right\Vert =1}\langle
\psi |\Omega |\psi \rangle \leq \sup_{\left\Vert \psi \right\Vert =1}\langle
\psi |\sum\limits_{\alpha }I_{B}\otimes M_{\alpha }|\psi \rangle \leq
\langle \psi |\psi \rangle =1.
\end{equation*}
$\square $

It follows that in the definition \ref{sep} it is sufficient to consider
sequences of operators, weakly convergent on a dense subspace of $\mathcal{H}
_{B}\otimes \mathcal{H}_{A}.$

Equipped the definition \ref{sep} and the construction of the integral (\ref
{sp}) below we can prove a rigorous version the corresponding result from
\cite{oppsw}.

\begin{proposition}
If the channel $\Phi $ is entanglement-breaking then its form is given by a
bounded operator $\Omega _{\Phi }\in \mathfrak{C}_{BA}.$If $\Phi $ has
representation (\ref{ebr}) then
\begin{equation}
\Omega _{\Phi }=\int\limits_{\mathcal{X}}\rho _{B}(x)\otimes \bar{M}_{A}(dx),
\label{sp}
\end{equation}
where the integral is defined in the proof.

Conversely, if a bounded operator $\Omega \in \mathfrak{C}_{BA}$ then $
\Omega =\Omega _{\Phi },$ where the channel $\Phi $ is entanglement-breaking.
\end{proposition}

\textit{Proof.} The proof of the first statement requires some theory of
integration with respect to a POVM.

Let $\mathcal{X}$ be a complete separable metric space with $\sigma -$
algebra of Borel subsets $\mathcal{B}$, let $\mathcal{H}$ is a separable
Hilbert space and $\left\{ M(E);B\in \mathcal{B}\right\} $ a POVM on $
\mathcal{X}$ with values in $\mathfrak{B}(\mathcal{H})$. Then for any $\psi
\in \mathcal{H}$ the set function $\left\{ \langle \psi |M(E)|\psi \rangle
;B\in \mathcal{B}\right\} $ is positive finite measure with total variation $
\left\Vert \psi \right\Vert ^{2}$; and for any $\psi ,\psi ^{\prime }\in
\mathcal{H}$ the set function $\left\{ \langle \psi |M(E)|\psi ^{\prime
}\rangle ;B\in \mathcal{B}\right\} $ is a complex measure of finite total
variation $\left\Vert \psi \right\Vert \left\Vert \psi ^{\prime }\right\Vert
$ as follows from the inequality
\begin{equation}
\left\vert \langle \psi |M(E)|\psi ^{\prime }\rangle \right\vert \leq \frac{
1 }{2}\left[ c\langle \psi |M(E)|\psi \rangle +c^{-1}\langle \psi |M(E)|\psi
^{\prime }\rangle \right] ;\quad c>0,  \label{inequa}
\end{equation}
due to positivity of the operator $M(E)$ for arbitrary Borel $E\subseteq
\mathcal{X}.$

A function $x\rightarrow \rho (x)$ with values in $\mathfrak{S}(\mathcal{H)}$
will be called Borel function if the scalar functions $x\rightarrow \langle
\psi |\rho (x)|\psi ^{\prime }\rangle $ are Borel for all $\psi ,\psi
^{\prime }\in \mathcal{H}.$ Let $\mu $ be a $\sigma -$ finite measure on $
\mathcal{X}$ then the function $x\rightarrow \rho (x)$ will be called
measurable if the functions $x\rightarrow \langle \psi |\rho (x)|\psi
^{\prime }\rangle $ are $\mu -$measurable for all $\psi ,\psi ^{\prime }\in
\mathcal{H}.$ This implies that for any $A\in \mathfrak{B}(\mathcal{H})$ the
scalar function $x\rightarrow \mathrm{Tr}\rho (x)A$ is $\mu -$measurable.
Since $\mathfrak{S}(\mathcal{H)}$ is separable with respect to the trace
norm distance, this implies, by theorem 3.5.3 from \cite{hp}, that $
x\rightarrow \rho (x)$ is strongly measurable in the sense that there is a
sequence of simple Borel functions $x\rightarrow \rho _{n}(x)$ such that $
\left\Vert \rho (x)-\rho _{n}(x)\right\Vert _{1}\rightarrow 0$ $(\mathrm{mod}
\mu )$. If $\mu $ is a probability measure then the Bochner integral $\int_{
\mathcal{X}}\rho (x)\mu (dx)=\lim_{n\rightarrow \infty }\int_{\mathcal{X}
}\rho _{n}(x)\mu (dx)$ exists and is an element of $\mathfrak{S}(\mathcal{H)}
.$

Now let $\Phi $ be entanglement-breaking channel with the
representation (\ref{ebr}). If $x\rightarrow \rho _{B}(x)$ is a
Borel function with values in
$\mathfrak{S}(\mathcal{H}_{B}\mathcal{)}$ and $M_{A}(dx)$ a POVM in
$ \mathcal{H}_{A}$ we define the integral
$\int\limits_{\mathcal{X}}\rho _{B}(x)\otimes \bar{M}_{A}(dx)\in
\mathfrak{B}(\mathcal{H}_{B}\otimes \mathcal{H}_{A})$ as follows.
For a simple function $\rho _{n}(x)=\sum\limits_{\alpha }\rho
_{\alpha }1_{E_{\alpha }}(x),$ where $ \left\{ E_{\alpha }\right\} $
is a finite decomposition of $\mathcal{X}$ , we put $\Omega
_{n}\equiv \int\limits_{\mathcal{X}}\rho _{n}(x)\otimes \bar{ M
}_{A}(dx)=\sum\limits_{\alpha }\rho _{\alpha }\otimes \bar{M}
_{A}(E_{\alpha }).$ Apparently $\Omega _{n}\in \mathfrak{C}_{BA},$
hence $ \left\Vert \Omega _{n}\right\Vert \leq 1.$ We also have
\begin{equation}
\langle \psi _{B}\otimes \psi _{A}\left\vert \Omega _{n}\right\vert \psi
_{B}^{\prime }\otimes \psi _{A}^{\prime }\rangle =\int\limits_{\mathcal{X}
}\langle \psi _{B}|\rho _{n}(x)|\psi _{B}^{\prime }\rangle \langle \bar{\psi}
_{A}^{\prime }|M_{A}(dx)|\bar{\psi}_{A}\rangle ,  \label{omegan}
\end{equation}
for all $\psi _{B},\psi _{B}^{\prime }\in \mathcal{H}_{B},$ $\psi _{A},\psi
_{A}^{\prime }\in \mathcal{H}_{A}.$ For two simple functions $\rho
_{n}(x),\rho _{m}^{\prime }(x)$ we have by the inequality (\ref{inequa})
\begin{equation*}
\left\vert \langle \psi _{B}\otimes \psi _{A}\left\vert \left( \Omega
_{n}-\Omega _{m}^{\prime }\right) \right\vert \psi _{B}^{\prime }\otimes
\psi _{A}^{\prime }\rangle \right\vert
\end{equation*}
\begin{equation*}
\leq \frac{1}{2}\left\Vert \psi _{B}\right\Vert \left\Vert \psi _{B}^{\prime
}\right\Vert \int\limits_{\mathcal{X}}\left\Vert \rho _{n}(x)-\rho
_{m}^{\prime }(x)\right\Vert _{1}\left[ c\langle \bar{\psi}_{A}|M_{A}(dx)|
\bar{\psi}_{A}\rangle +c^{-1}\langle \bar{\psi}_{A}^{\prime }|M_{A}(dx)|\bar{
\psi}_{A}^{\prime }\rangle \right] .
\end{equation*}
Take $\left\{ \rho _{n}(x)\right\} $ such that $\left\Vert \rho (x)-\rho
_{n}(x)\right\Vert _{1}\rightarrow 0$ $(\mathrm{mod}\mu _{\sigma })\ $ where
$\sigma $ is a state in $\mathfrak{S}(\mathcal{H}_{A})$ of full rank, then
it follows that the forms $\langle \psi _{BA}\left\vert \Omega
_{n}\right\vert \psi _{BA}^{\prime }\rangle $ converge on a dense subspace
of $\mathcal{H}_{B}\otimes \mathcal{H}_{A}$ and are uniformly bounded. Hence
there exists uniquely defined bounded operator $\Omega ,$ which is the limit
of $\Omega _{n}$ in the weak operator topology. We define $\int\limits_{
\mathcal{X}}\rho _{B}(x)\otimes \bar{M}_{A}(dx)=\Omega ,$ and it follows
from the definition that $\Omega \in \mathfrak{C}_{BA}.$

From the definition (\ref{relation}) of the form $\Omega _{\Phi }$ we obtain
\begin{equation*}
\Omega _{\Phi }(\psi _{B}\otimes \psi _{A};\psi _{B}^{\prime }\otimes \psi
_{A}^{\prime }\rangle =\int\limits_{\mathcal{X}}\langle \psi _{B}|\rho
_{B}(x)|\psi _{B}^{\prime }\rangle \langle \bar{\psi}_{A}^{\prime
}|M_{A}(dx)|\bar{\psi}_{A}\rangle ,
\end{equation*}
which is the limit of (\ref{omegan}). Since, on the other hand, (\ref{omegan}
) converge to the form defined by the operator $\Omega ,$ we obtain that the
form $\Omega _{\Phi }$ is defined by the operator $\Omega .$

Conversely, let a bounded operator $\Omega \in \mathfrak{C}_{BA}$. Fix a
state $\sigma $ in $\mathfrak{S}(\mathcal{H}_{A})$ of full rank. Using the
fact that weak convergence of operators $\Omega =\sum\limits_{\alpha }\rho
_{\alpha }\otimes M_{\alpha }\in \mathfrak{C}_{BA}$ implies weak convergence
of density operators $\sigma _{BA}=\sum\limits_{\alpha }\rho _{\alpha
}\otimes \sigma ^{1/2}M_{\alpha }\sigma ^{1/2}$, we can prove that the state
$\sigma _{BA}=\left( I_{B}\otimes \sigma ^{1/2}\right) \Omega \left(
I_{B}\otimes \sigma ^{1/2}\right) $ is separable. Basing on the
decomposition (\ref{sepa}) for $\sigma $ and arguing as in \cite{hsw} we can
construct POVM $M_{A}(dx)$ and the family of states $\rho _{B}(x)$ such that
$\Omega $ is given by (\ref{sp}) and hence $\Omega =\Omega _{\Phi }$ for the
corresponding entanglement-breaking channel $\Phi $ given by (\ref{ebr}). $
\square $

\section{Bosonic Gaussian channels}

Let $(Z,\Delta )$ be a coordinate symplectic space, $\dim Z=2s,$ with the
symplectic form
\begin{equation*}
\Delta (z,z^{\prime })=z^{t}\Delta z^{\prime };\quad \Delta =\mathrm{diag}
\left[
\begin{array}{cc}
0 & -1 \\
1 & 0
\end{array}
\right] _{j=1,\dots ,s}
\end{equation*}
where $^{t}$ denotes transposition in $Z.$ Let $\mathcal{H}$ be the space of
irreducible representation of the Canonical Commutation Relations (CCR)
\begin{equation}
W(z)W(z^{\prime })=\exp \left( \frac{i}{2}\Delta (z,z^{\prime })\right)
W(z+z^{\prime }).  \label{ccr}
\end{equation}
Here $W(z)=\exp (iR\cdot z);\,z\in Z$, is the Weyl system and $
R=[q_{1},p_{1};\dots ,q_{s},p_{s}]$ is the row-vector of the canonical
variables in $\mathcal{H}$, see \cite{qi} for more detail. We will continue
to denote $^{\top }$ transposition in $\mathcal{H}$ associated to a basis $
\left\{ |i\rangle \right\} $; from (\ref{ccr}) it follows that $W(z)^{\top
}=\exp (iR^{\top }\cdot z);\,z\in Z,$ is the Weyl system for the symplectic
space $(Z,-\Delta ).$ The \textit{canonical transposition} associated with
the Fock basis in $\mathcal{H}$ is given by $R^{\top }=[q_{1},-p_{1};\dots
,q_{s},-p_{s}];$ all the others are obtained from this by unitary
conjugations. In what follows transposition is arbitrary if not stated
otherwise.

Let $\mathcal{H}_{A}$ be the representation space of CCR with a coordinate
symplectic space $(Z_{A},\Delta _{A})$ and the Weyl system $W_{A}(\cdot ),$
with a similar description for $\mathcal{H}_{B}.$ Let $\Phi :\mathfrak{T}(
\mathcal{H}_{A})\mapsto \mathfrak{T}(\mathcal{H}_{B})$ be a channel. Then
the following relation holds
\begin{equation}
\Omega _{\Phi }=\frac{1}{(2\pi )^{s}}\int_{Z_{B}}W_{B}(-z)\otimes \Phi
^{\ast }\left( W_{B}(z)\right) ^{\top }d^{2s}z,  \label{invers}
\end{equation}
in the sense of forms defined on $\mathcal{H}_{B}\times \mathcal{H}_{A}.$
Here $d^{2s}z$ is the element of symplectic volume in $Z_{B}$ and $^{\top }$
is a transposition in $\mathcal{H}_{A}.$

Indeed,
\begin{equation}
\langle \bar{\psi}^{\prime }|\Phi ^{\ast }\left( W_{B}(z)\right) |\bar{\psi}
\rangle =\mathrm{Tr}\Phi (|\bar{\psi}\rangle \langle \bar{\psi}^{\prime
}|)W_{B}(z),  \label{eq}
\end{equation}
where $\Phi (|\bar{\psi}\rangle \langle \bar{\psi}^{\prime }|)$ is
trace-class operator and hence (\ref{eq}) is square-integrable as a function
of $z\in Z_{B}$. Similarly, the function $\langle \psi |W(-z)|\psi ^{\prime
}\rangle $ is also square-integrable. By the inversion formula for the
noncommutative Fourier transform (cf. ch. V of \cite{qi}), applied to the
right-hand side of (\ref{eq}),
\begin{eqnarray*}
\Omega _{\Phi }\left( \psi _{B}\otimes \psi _{A};\psi _{B}^{\prime }\otimes
\psi _{A}^{\prime }\right)  &=&\langle \psi _{B}|\Phi (|\bar{\psi}
_{A}\rangle \langle \bar{\psi}_{A}^{\prime }|)|\psi _{B}^{\prime }\rangle  \\
&=&\frac{1}{(2\pi )^{s}}\int \langle \psi _{B}|W_{B}(-z)|\psi _{B}^{\prime
}\rangle \langle \bar{\psi}_{A}^{\prime }|\Phi ^{\ast }\left(
W_{B}(z)\right) |\bar{\psi}_{A}\rangle d^{2s}z,
\end{eqnarray*}
so that finally we get the formula (\ref{invers}).

\bigskip Now let $\Phi $ be a (centered) Gaussian channel \cite{cegh},
\begin{equation*}
\Phi ^{\ast }\left( W_{B}(z)\right) =W_{A}(Kz)\exp \left( -\frac{1}{2}
z^{t}\mu z\right) ,\quad z\in Z_{B},
\end{equation*}
where
\begin{equation}
\mu \geq \pm \frac{i}{2}\Delta _{K};\quad \Delta _{K}=\Delta
_{B}-K^{t}\Delta _{A}K.  \label{nid}
\end{equation}
Then (\ref{invers}) implies
\begin{equation}
\Omega _{\Phi }=\frac{1}{(2\pi )^{s}}\int W_{B}(-z)\otimes W_{A}(Kz)^{\top
}\exp \left( -\frac{1}{2}z^{t}\mu z\right) d^{2s}z.  \label{inv}
\end{equation}
The unitary operators $W_{BA}(z)=W_{B}(z)\otimes W_{A}(-Kz)^{\top }=\exp
\left( iR_{BA}\cdot z\right) ,$ where
\begin{equation*}
R_{BA}=R_{B}\otimes I-I\otimes R_{A}^{\top }K,
\end{equation*}
satisfy the Weyl-Segal CCR with (possibly degenerate) symplectic
form determined by the matrix $\Delta _{K}.$ This representation in
the space $ \mathcal{H}_{B}\otimes \mathcal{H}_{A}$ is reducible
(even if $\Delta _{K}$ is nondegenerate, see the footnote below).
Finally
\begin{equation}
\Omega _{\Phi }=\frac{1}{(2\pi )^{s}}\int_{Z_{B}}\exp \left( -iR_{BA}\cdot
z\right) \exp \left( -\frac{1}{2}z^{t}\mu z\right) d^{2s}z.  \label{inv1}
\end{equation}

\textit{If }$\mu $\textit{\ is nondegenerate, then the form }$\Omega
_{\Phi } $\textit{\ is given by bounded operator with}
\begin{equation}
\left\Vert \Omega _{\Phi }\right\Vert \leq \frac{1}{\sqrt{\det \mu }}.
\label{bound2}
\end{equation}
This follows from (\ref{inv1}) taking into account the fact that norm of the
Weyl operators is equal to 1.

In what follows we shall consider the four basic cases. Later it
will be convenient to use reverse enumeration, so we start with the
last, the most degenerate case.

\textbf{Case 4.} If $\mu =0\ $then $\Delta _{K}=0$ and $R_{BA}$ is the
vector operator with commuting selfadjoint components. Thus
\begin{equation}
\Omega _{\Phi }=\frac{1}{(2\pi )^{s}}\int \exp \left( -iR_{BA}\cdot z\right)
d^{2s}z=\delta \left( R_{BA}\right) ,  \label{delta}
\end{equation}
where $\delta \left( \cdot \right) $ is Dirac's delta-function. In this case
$\Omega _{\Phi }$ is a nonclosable form. Note that for the ideal channel $
(A=B)$ this gives
\begin{equation}
|\Omega \rangle \langle \Omega |=\frac{1}{(2\pi )^{s}}\int W_{B}(-z)\otimes
W_{A}(z)^{\top }d^{2s}z=\delta \left( R_{BA}\right) ,  \label{ideal}
\end{equation}
where $R_{BA}=R_{B}\otimes I-I\otimes R_{A}^{\top }.$

\textbf{Case 3.} If $\mu >0,\Delta _{K}=0$ then again $R_{BA}=R_{B}\otimes
I-I\otimes R_{A}^{\top }K$ is the vector operator with commuting selfadjoint
components and the integral (\ref{inv1}) is just the multivariate Gaussian
density as a function of $R_{BA}:$
\begin{equation*}
\Omega _{\Phi }=\frac{1}{\sqrt{\det \mu }}\exp \left( -\frac{1}{2}R_{BA}\mu
^{-1}R_{BA}^{t}\right) .
\end{equation*}
Since the spectrum of $R_{BA}$ contains $0,$ we have $\left\Vert \Omega
_{\Phi }\right\Vert =\frac{1}{\sqrt{\det \mu }}.$

\begin{proposition}
Assume that $\Delta _{K}$ is nondegenerate, then
\begin{equation}
\left\Vert \Omega _{\Phi }\right\Vert =\frac{1}{\sqrt{\det \Delta _{K}\det
\left[ \mathrm{abs}\left( \Delta _{K}^{-1}\mu \right) +I/2\right] }}.
\label{bound1}
\end{equation}

\textbf{Case 1.} If $\mu -\frac{i}{2}\Delta _{K}$ is nondegenerate,
then
\begin{equation}
\Omega _{\Phi }=\frac{1}{\sqrt{\det \left( \mu -\frac{i}{2}\Delta
_{K}\right) }}\exp \left( -R_{BA}\epsilon R_{BA}^{t}\right) ,  \label{omgaus}
\end{equation}
where $\epsilon =\mathrm{arccot}\left( 2\Delta _{K}^{-1}\mu \right) \Delta
_{K}^{-1}.$

\textbf{Case 2.} If $\mu -\frac{i}{2}\Delta _{K}$ is maximally
degenerate i.e. $\mbox{rank}\left( \mu -\frac{i}{2}\Delta
_{K}\right) =s$, then $\Omega _{\Phi }=\frac{1}{\sqrt{\det \Delta
_{K}}}P_{0},$ where $P_{0}$ is the projection onto the kernel of
positive selfadjoint operator $R_{BA}\mu ^{-1}R_{BA}^{t}-sI$ and
$\left\Vert \Omega _{\Phi }\right\Vert =\frac{1}{ \sqrt{\det \Delta
_{K}}}.$
\end{proposition}

\textit{Proof.} If $\Delta _{K}$ is nondegenerate then $\mu $ is
also nondegenerate, so $\Omega _{\Phi }$ is given by bounded
operator. Rewriting (\ref{inv}) and again using the inversion
formula we get
\begin{equation*}
\Omega _{\Phi }=\frac{1}{(2\pi )^{s}\sqrt{\det \Delta _{K}}}\int \exp \left(
-iR_{BA}\cdot z\right) \exp \left( -\frac{1}{2}z^{t}\mu z\right) d_{\Delta
_{K}}^{2s}z=\frac{1}{\sqrt{\det \Delta _{K}}}\rho _{K},
\end{equation*}
where $d_{\Delta _{K}}^{2s}z=$ $\sqrt{\det \Delta _{K}}d^{2s}z$ is
the volume element corresponding to the symplectic form $z^{t}\Delta
_{K}z^{\prime },$ and $\rho _{K}$ has the expression in the
canonical variables $R_{BA}$ as the Gaussian density
operator\footnote{ Notice that $\rho _{K}$ is not a proper density
operator in the space $ \mathcal{H}_B\otimes \mathcal{H}_A$ since
$R_{BA}$ generate a reducible representation $W_{BA}(z)$ of CCR in
that space. Actually $\rho _{K}$ is tensor product of the Gaussian
density operator in the space where $ W_{BA}(z) $ act irreducibly
with the identity in the complementary space, reflecting the
multiplicity of the representation.} with zero mean and the
covariance matrix $\mu .$ The value (\ref{bound1}) is just the
maximal eigenvalue of this Gaussian density operator, multiplied by
$\left( \det \Delta _{K}\right) ^{-1/2}.$ There is a nondegenerate
transformation $T$ such that
\begin{equation}
\tilde{\mu}=T^{t}\mu T=\mathrm{diag}\left[
\begin{array}{cc}
\mu _{j} & 0 \\
0 & \mu _{j}
\end{array}
\right] ,\quad \Delta =T^{t}\Delta _{K}T=\mathrm{diag}\left[
\begin{array}{cc}
0 & -1 \\
1 & 0
\end{array}
\right] ,  \label{one-mode-alpha}
\end{equation}
where $\mu _{j}\geq \frac{1}{2},\quad j=1,\dots ,s$ (see the next Section).
Then $\Delta ^{-1}\tilde{\mu}=\mathrm{diag}\left[
\begin{array}{cc}
0 & \mu _{j} \\
-\mu _{j} & 0
\end{array}
\right] $ and $\Delta _{K}^{-1}\mu =T\Delta ^{-1}\tilde{\mu}T^{-1}$ is
matrix of the operator with eigenvalues $\pm i\mu _{j},$ so that the
operator $\mathrm{abs}\left( \Delta _{K}^{-1}\mu \right) =T\mathrm{diag}
\left[
\begin{array}{cc}
\mu _{j} & 0 \\
0 & \mu _{j}
\end{array}
\right] T^{-1}$ has the eigenvalues $\mu _{j}$ of multiplicity 2 . The
operator $\rho _{K}$ splits into the normal modes decomposition

\begin{equation}
\rho _{K}=\bigotimes_{j=1}^{s}\rho ^{(j)},  \label{charct. fct.}
\end{equation}
with $\rho ^{(j)}$ being the elementary one-mode Gaussian density operator

\begin{equation}
\rho ^{(j)}=\frac{1}{\mu _{j}+\frac{1}{2}}\left( \frac{\mu _{j}-\frac{1}{2}}{
\mu _{j}+\frac{1}{2}}\right) ^{\tilde{n}_{j}},  \label{one-mode}
\end{equation}
where $\tilde{n}_{j}=\frac{1}{2}\left( \tilde{q}_{j}^{2}+\tilde{p}
_{j}^{2}-1\right) $ is the number operator for the $j-$th mode (see ch. V of
\cite{qi}). Here the new canonical variables $\tilde{R}=[\tilde{q}_{1},\dots
,\tilde{p}_{s}]$ are related to the old ones by the formula $\tilde{R}
=R_{BA}T$. The maximal eigenvalue of $\rho _{K}$ is thus equal to
\begin{equation*}
\prod\limits_{j=1}^{s}\frac{1}{\mu _{j}+\frac{1}{2}}=\frac{1}{\sqrt{\det
\left[ \mathrm{abs}\left( \Delta _{K}^{-1}\mu \right) +I/2\right] }}.
\end{equation*}

Since $\mu -\frac{i}{2}\Delta _{K}=\Delta _{K}\left( \Delta
_{K}^{-1}\mu - \frac{i}{2}\right) ,$ the condition that $\mu
-\frac{i}{2}\Delta _{K}$ is nondegenerate (Case 1) is equivalent to
$\mu _{j}>\frac{1}{2},\quad j=1,\dots ,s,$ i.e. the decomposition
(\ref{charct. fct.}) has no pure component. Coming back from
(\ref{charct. fct.}), (\ref{one-mode}) to the initial canonical
observables $R_{BA}$ gives
\begin{equation}
\rho _{K}=c\exp \left( -R_{BA}\epsilon R_{BA}^{t}\right) ,  \label{Gibbs}
\end{equation}
where
\begin{equation}
c=\prod\limits_{j=1}^{s}\frac{1}{\sqrt{\mu _{j}^{2}-\frac{1}{4}}}=\frac{1}{
\sqrt{\det \left( \Delta _{K}^{-1}\mu -\frac{i}{2}\right) }}  \label{consta}
\end{equation}
and $\epsilon $ is found from
\begin{equation}
2\Delta _{K}^{-1}\mu =\cot \epsilon \Delta _{K},  \label{cot}
\end{equation}
whence the formula (\ref{omgaus}).

The case 2 corresponds to $\mu _{j}=\frac{1}{2},$ $j=1,\dots ,s$. Then $\rho
_{K}$ is the projection onto the kernel of the positive selfadjoint operator
\begin{equation*}
2\sum\limits_{j=1}^{s}\tilde{n}_{j}=\sum\limits_{j=1}^{s}\left( \tilde{q}
_{j}^{2}+\tilde{p}_{j}^{2}-1\right) =\tilde{R}\tilde{R}^{t}-sI=R_{BA}\mu
^{-1}R_{BA}^{t}-sI,
\end{equation*}
so that
\begin{equation*}
\Omega _{\Phi }=\frac{1}{\sqrt{\det \Delta _{K}}}\delta _{0}\left( R_{BA}\mu
^{-1}R_{BA}^{t}-sI\right)
\end{equation*}
(Kronecker's delta), and $\left\Vert \Omega _{\Phi }\right\Vert =\frac{1}{
\sqrt{\det \Delta _{K}}}.\square $

\textbf{Example.} Consider attenuator/amplifier in one mode,
\begin{equation*}
\Delta =\left[
\begin{array}{cc}
0 & -1 \\
1 & 0
\end{array}
\right]
\end{equation*}
given by
\begin{equation*}
\Phi ^{\ast }\left( W(z)\right) =W(kz)\exp \left( -\frac{m}{2}\left\vert
z\right\vert ^{2}\right) ,\quad z\in \mathbb{R}^{2}.
\end{equation*}
where $k,m\geq 0.$ Then (\ref{nid}) reduces to $m\geq \frac{\left\vert
k^{2}-1\right\vert }{2}$ and $\det \Delta _{K}=\left( k^{2}-1\right) ^{2}.$
The relation (\ref{bound1}) gives
\begin{equation*}
\left\Vert \Omega _{\Phi }\right\Vert =\frac{1}{m+\frac{\left\vert
k^{2}-1\right\vert }{2}}.
\end{equation*}
The channel is entanglement-breaking if and only if $m\geq \frac{k^{2}+1}{2}$
i.e. $m+\frac{\left\vert k^{2}-1\right\vert }{2}\geq \max \left\{
1,k^{2}\right\} $ \cite{h} which agrees with Lemma 1. Also if $1-\frac{
\left\vert k^{2}-1\right\vert }{2}\leq m<\frac{k^{2}+1}{2}$ then the channel
is not entanglement-breaking while still $\left\Vert \Omega _{\Phi
}\right\Vert \leq 1.$ Together with $m\geq \frac{\left\vert
k^{2}-1\right\vert }{2}$ this gives $k>1$ (amplifier) and the lower bound $
m\geq \max \left\{ \frac{3-k^{2}}{2},\frac{k^{2}-1}{2}\right\} .$

Assuming the canonical transposition $^{\top }$ we have $[q;p]^{\top
}=[q;-p].$ Then, using the relation (\ref{one-mode}), one can obtain in the
case 1 $\left( m>\frac{\left\vert k^{2}-1\right\vert }{2}\right) $
\begin{equation}
\Omega _{\Phi }=\frac{1}{\sqrt{m^{2}-\frac{\left( k^{2}-1\right) ^{2}}{4}}}
\exp \left\{ -\frac{1}{2\left\vert k^{2}-1\right\vert }\ln \frac{m+\frac{
\left\vert k^{2}-1\right\vert }{2}}{m-\frac{\left\vert k^{2}-1\right\vert }{
2 }}\left[ \left( q_{B}-kq_{A}\right) ^{2}+\left( p_{B}+kp_{A}\right) ^{2}
\right] \right\} .  \label{1mode}
\end{equation}
The case 2 corresponds to $m=\frac{\left\vert k^{2}-1\right\vert }{2};$ then
one obtains $\Omega _{\Phi }=\left\vert k^{2}-1\right\vert ^{-1}P_{0},$
where $P_{0}$ is the projection onto the eigenspace of the operator $\left(
q_{B}-kq_{A}\right) ^{2}+\left( p_{B}+kp_{A}\right) ^{2},$ corresponding to
its lowest eigenvalue $\left\vert k^{2}-1\right\vert .$

In general, the decomposition (\ref{charct. fct.}) means that in the case 1
the operator $\Omega _{\Phi }$ can be decomposed into tensor product of
operators of the form (\ref{1mode}), and similarly in the case 2.

\section{A decomposition of the Gaussian CJ form}

Recall that $2s=\dim Z_{B}$ and denote by $r_{\alpha }=\mbox{rank}\,\alpha $
-- the rank of a $2s\times 2s-$matrix $\alpha .$ The following result is a
generalization of the Williamson's lemma (cf. \cite{cegh1}).

\begin{lemma}
\label{lem} Let $\mu $ be a real symmetric matrix, $\Delta _{K}$ --
a real skew-symmetric matrix such that $\mu -\frac{i}{2}\Delta
_{K}\geq 0.$ Then there is a nondegenerate matrix $T$ such that
\begin{eqnarray}
T^{t}\mu T&=&\left[
\begin{array}{ccc}
\tilde{\mu} & 0 & 0 \\
0 & I/2 & 0 \\
0 & 0 & 0
\end{array}
\right]
\begin{array}{l}
\}r_{\Delta _{K}} \\
\}r_{\mu }-r_{\Delta _{K}} \\
\}2s-r_{\mu }
\end{array}
,\quad  \label{decom1} \\
T^{t}\Delta _{K} T &=&\left[
\begin{array}{ccc}
\Delta & 0 & 0 \\
0 & 0 & 0 \\
0 & 0 & 0
\end{array}
\right],  \label{decom2}
\end{eqnarray}
where \
\begin{equation*}
\Delta =\mathrm{diag}\left[
\begin{array}{cc}
0 & -1 \\
1 & 0
\end{array}
\right] _{j=1,\dots ,r_{\Delta _{K}}/2},\quad \tilde{\mu}=\mathrm{diag}\left[
\begin{array}{cc}
\mu _{j} & 0 \\
0 & \mu _{j}
\end{array}
\right] _{j=1,\dots ,r_{\Delta _{K}}/2},
\end{equation*}
and $\mu _{j}\geq 1/2.$
\end{lemma}

Notice that $r_{\mu }$ can be odd. Denote $d_{3}=r_{\mu }-r_{\Delta
_{K}},d_{4}=2s-r_{\mu }$ the dimensionalities of the last two blocks in the
decompositions (\ref{decom1}), (\ref{decom2}). Let us further arrange the
block diagonal matrix $\tilde{\mu}$ as
\begin{equation*}
\tilde{\mu}=\left[
\begin{array}{cc}
\tilde{\mu}^{(1)} & 0 \\
0 & \tilde{\mu}^{(2)}
\end{array}
\right]
\begin{array}{c}
\}d_{1} \\
\}d_{2}
\end{array}
\end{equation*}
by putting first the blocks with $\mu _{j}>1/2$ and then -- the blocks with $
\mu _{j}=1/2.$ We have $r_{\Delta _{K}}=d_{1}+d_{2},r_{\mu -\frac{i}{2}
\Delta _{K}}=d_{1}+d_{2}/2+d_{3},$ whence
\begin{equation*}
d_{1}=r_{\Delta _{K}}-2(r_{\mu }-r_{\mu -\frac{i}{2}\Delta
_{K}}),d_{2}=2(r_{\mu }-r_{\mu -\frac{i}{2} \Delta _{K}}).
\end{equation*}
Let $\tilde{e}_{j}=T^{-1}e_{j};j=1,\dots ,2s$ be the basis in $ Z_{B}$ in
which $\mu ,\Delta _{K}$ have the block diagonal form (\ref{decom1}), (\ref
{decom2}) and let $\tilde{Z}_{k}$ be the $d_{k}-$dimensional subspace
spanned the vectors $\tilde{e}_{j}$ corresponding to the $k-$th block in the
decompositions, $k=1,\dots ,4.$ Then we have the direct sum decomposition
\begin{equation}
Z_{B}=\tilde{Z}_{1}+\tilde{Z}_{2}+\tilde{Z}_{3}+\tilde{Z}_{4}  \label{dsum}
\end{equation}

\bigskip By making the substitution $T^{-1}z=\tilde{z}$ in (\ref{inv1}), we
have $ \tilde{z}=\left[ \tilde{z}_{1},\tilde{z}_{2},\tilde{z}_{3},\tilde{z}
_{4} \right] ^{t}$ and
\begin{equation*}
\Omega _{\Phi }=\frac{1}{(2\pi )^{s}\left\vert \det T\right\vert }\int \int
\int \int \exp \sum_{k=1}^{4}\left( -iR_{BA}T\tilde{z}_{k}-\frac{1}{2}
\tilde{z}_{k}^{t}\tilde{\mu}^{(k)}\tilde{z}_{k}\right) d\tilde{z}_{1}d\tilde{
z}_{2}d\tilde{z}_{3}d\tilde{z}_{4},
\end{equation*}
where $\tilde{\mu}_{(3)}=I_{d_{3}}/2,$ $\tilde{\mu}_{(4)}=0_{d_{4}}$ and the
components of $R_{BA}T\tilde{z}_{k}$ and $R_{BA}T\tilde{z}_{l}$ commute for $
k\neq l$ by (\ref{decom2}). Hence the exponent under the integral splits
into product of four mutually commuting exponents, and the CJ form $\Omega
_{\Phi }$ can be decomposed into the product of commuting expressions of the
types considered in the cases 1-4 above (with possibly odd dimensionalities
for $\tilde{z}_{1},\tilde{z}_{2}$):
\begin{equation}
\Omega _{\Phi }=\frac{\left\vert \det T\right\vert }{(2\pi )^{s}}
\prod_{k=1}^{4}\int_{\tilde{Z}_{k}}\exp \left( -iR_{BA}T\tilde{z}_{k}- \frac{
1}{2}\tilde{z}_{k}^{t}\tilde{\mu}_{(k)}\tilde{z}_{k}\right) d\tilde{z} _{k}.
\label{prod}
\end{equation}

This product can be further transformed into tensor product in the space $
\mathcal{H}_{B}\otimes \mathcal{H}_{A}$ as follows. Consider the direct sum $
Z_{B}+Z_{A}$ equipped with the symplectic form defined by the skew-symmetric
matrix
\begin{equation*}
\Delta _{AB}=\left[
\begin{array}{cc}
I & K^{t} \\
0 & I
\end{array}
\right] \left[
\begin{array}{cc}
\Delta _{B} & 0 \\
0 & -\Delta _{A}
\end{array}
\right] \left[
\begin{array}{cc}
I & 0 \\
K & I
\end{array}
\right] =\left[
\begin{array}{cc}
\Delta _{K} & -K^{t}\Delta _{A} \\
-\Delta _{A}K & -\Delta _{A}
\end{array}
\right] .
\end{equation*}
This form is nondegenerate since it is determined by the product of
nondegenerate matrices; moreover, its restriction to $Z_{B}$
coincides with $ \Delta _{K}$. The unitary operators
\begin{equation*}
W(z_{B},z_{A})=\exp i\left( R_{BA}z_{B}-I\otimes R_{A}^{\top }z_{A}\right)
=\exp i\left( [R_{B},-R_{A}^{\top }]\left[
\begin{array}{cc}
I & 0 \\
K & I
\end{array}
\right] \left[
\begin{array}{c}
z_{B} \\
z_{A}
\end{array}
\right] \right)
\end{equation*}
give an irreducible representation of CCR on the symplectic space $\left(
Z_{B}+Z_{A},\Delta _{AB}\right) ,$ and at the same time $W(z_{B},0)=\exp
i\left( R_{BA}z_{B}\right)=W_B(z_{B}).$ It follows that the direct sum
decomposition (\ref{dsum}) can be extended to the decomposition
\begin{equation}
Z_{B}+Z_{A}=\tilde{Z}_{1}\oplus \tilde{Z}_{2}\oplus \tilde{Z}^{\prime}_{3}
\oplus \tilde{Z}^{\prime}_{4} \oplus Z_{0},  \label{decom3}
\end{equation}
where $\tilde{Z}^{\prime}_{3}\supseteq\tilde{Z}_{3}, \tilde{Z}
^{\prime}_{4}\supseteq\tilde{Z}_{4}$ and $\oplus $ means that
$\Delta _{AB}$ is nondegenerate on each of the five subspaces
(provided they are nontrivial) and zero between the different
subspaces.

To prove this, we use the fact: if $\left( Z,\Delta \right) $ is a
symplectic space and $\tilde{Z}$ a subspace of $Z$ such that the
restriction $\Delta |_{\tilde{Z}}$ is nondegenerate, then there is a
unique subspace $ \tilde{Z}^{\perp }$ such that $Z=\tilde{Z}\oplus
\tilde{Z}^{\perp }.$ From the decomposition (\ref{decom2}) and the
fact that $\Delta _{AB}|_{Z_{B}}=\Delta _{K}$ it then follows that
$Z_{B}+Z_{A}=\tilde{Z} _{1}\oplus \tilde{Z}_{2}\oplus M,$ where
$M\supseteq\tilde{Z}_{3}+\tilde{Z} _{4}.$ Then $\left( M,\Delta
_{AB}|_{M}\right) $ is itself a symplectic space and $\Delta
_{AB}|_{M}$ is identically zero on $\tilde{Z} _{3}+\tilde{Z }_{4}.$
The basis $\left\{ \tilde{e}_{j}\right\} $ in $\tilde{Z}
_{3}+\tilde{ Z}_{4}$ can be complemented by the system $\left\{
h_{j}\right\} $ in $M$ such that $\Delta
_{AB}(\tilde{e}_{j},h_{k})=\delta _{jk},$ $\Delta _{AB}(
\tilde{e}_{j},\tilde{e}_{k})=\Delta _{AB}(h_{j},h_{k})=0.$ Let the
subspaces $\tilde{Z}^{\prime}_{3},\tilde{Z}^{\prime}_{4}\subseteq M$
be spanned by the vectors $\tilde{e}_{j},h_{j}$ such that the
corresponding $\tilde{e}_{j}$ span, respectively, $\tilde{Z}
_{3},\tilde{Z}_{4}.$ Then by construction $ \Delta _{AB}$ is
nondegenerate on $\tilde{Z}^{\prime}_{3},\tilde{Z} ^{\prime}_{4}$
and equals zero between $\tilde{Z }_{1},\tilde{Z}_{2},\tilde{
Z}^{\prime}_{3},\tilde{Z}^{\prime}_{4}.$ Denoting $ Z_{0}=\left[
\tilde{Z} _{1}\oplus \tilde{Z}_{2}\oplus \tilde{Z}^{\prime}_{3}
\oplus \tilde{Z} ^{\prime}_{4}\right] ^{\perp },$ we obtain
(\ref{decom3}).

But this decomposition means that $\mathcal{H}_{B}\otimes \mathcal{H}_{A}$
can be splits into tensor product $\mathcal{H}_{1}\otimes \mathcal{H}
_{2}\otimes \mathcal{H}_{3}\otimes \mathcal{H}_{4}\otimes \mathcal{H}_{0}$
such that $\exp \left( iR_{BA}T\tilde{z}_{k}\right) $ acts nontrivially in $
\mathcal{H}_{k}$ for $k=1,\dots ,4\ $so that the product (\ref{prod}) can be
transformed into the tensor product.

Since in the cases 1-3 the integrals in the product are given by bounded
operators, we can complement (\ref{bound2}) as follows:

\begin{proposition}
Nondegeneracy of the matrix $\mu $ is necessary and sufficient for the form $
\Omega _{\Phi }$ to be defined by a bounded operator.
\end{proposition}

Finally, let us give interpretation of the decomposition (\ref{dsum}) in
terms of open system dynamics. Then we have the composite system $
Z_{A}\oplus Z_{D}=Z_{B}\oplus Z_{E}$ (input+noise
(distortion)=output+environment) which evolves reversibly according to the
unitary operator $U$ in the space $\mathcal{H}_{A}\oplus \mathcal{H}_{D}=
\mathcal{H}_{B}\oplus \mathcal{H }_{E}.$ The dynamical equations of the
channel in the Heisenberg picture can be written as
\begin{equation}
R_{B}^{\prime }\equiv U^{\ast }\left( R_{B}\otimes I_{E}\right)
U=R_{A}K\otimes I_{D}+I\otimes R_{D},  \label{chan}
\end{equation}
where $R_{D}$ is the vector of noise variables having the commutator
matrix $ \Delta _{K}=\Delta _{B}-K^{t}\Delta _{A}K$ and the
covariance matrix $\mu .$ The lemma \ref{lem} implies that $R
_{D}=[\tilde{R}_{D}^{q},\tilde{R} _{D}^{c}]T^{-1}$ where
$\tilde{R}_{D}^{q}$ is the $d_{1}+d_{2}=r_{\Delta _{K}}-$
dimensional subvector of quantum noise canonical observables with
the commutator matrix $\Delta $ and the covariance matrix
$\tilde{\mu}$ and $ \tilde{R}_{D}^{c}$ is $d_{3}+d_{4}=2s-r_{\Delta
_{K}}-$ dimensional subvector of commuting classical noise variables
(which commute also with $ \tilde{R}_{E}^{q}$). The summands in the
decomposition (\ref{dsum}) correspond to 1) quantum noise
observables in the nondegenerate Gaussian state, 2) quantum noise
observables in the pure Gaussian state, 3) classical noise variables
with positive variance, 4) trivial classical noise variables with
zero variance.

\bigskip

\textbf{Acknowledgments.} The author is grateful to V. Giovannetti and M. E.
Shirokov for discussions. The first version of this work was partially
supported by RFBR grant 09-01-00424 and the RAS program ``Mathematical
control theory'' . The new improved and extended version was written during
the author's stay at the Mittag-Leffler Institute in the frame of the
program ``Quantum Information Theory'', Fall 2010.

\end{document}